\documentclass[prb,twocolumn,showpacs]{revtex4}
\usepackage{bm}
\usepackage{graphicx}
\usepackage{amssymb}
\usepackage{amsmath}
\usepackage{color}
\newcommand{\nix}[1]{}

\begin{document}
\title{
Diluted magnetic semiconductor heterostructure AlSb/InAs/ZnMnTe 
with \\ giant Zeeman effect for two dimensional electrons in InAs
}
\author{Ya. V. Terent'ev,$^{1,2}$ C. Zoth,$^{1}$   V. V. Bel'kov,$^{1,2}$  P.
Olbrich,$^{1}$ C. Drexler,$^{1}$ V. Lechner,$^{1}$ P.~Lutz,$^{1}$ A. N. Semenov,$^{2}$ V. A.
Solov'ev,$^{2}$ I. V. Sedova,$^{2}$ G. V. Klimko,$^{2}$ T. A. Komissarova,$^{2}$ S. V. Ivanov,$^{2}$
and S. D. Ganichev$^{1}$}
\affiliation{$^1$ Terahertz Center, University of Regensburg, 93040 Regensburg, Germany}
\affiliation{$^2$ Ioffe Physical-Technical Institute, 
194021 St.\,Petersburg, Russia}
\begin{abstract}
A new approach to the growth of diluted magnetic semiconductors 
with two dimensional electron gas 
in InAs quantum well 
has been developed. The method is
based on molecular-beam epitaxy of coherent ``hybrid'' AlSb/InAs/(Zn,Mn)Te
heterostructures with a III-V/II-VI 
interface inside. The
giant Zeeman splitting of the InAs conduction band 
caused by exchange interaction with Mn$^{2+}$ ions has
been proved by measuring the microwave radiation induced spin polarized
electric currents.
\end{abstract}


\date{\today}

\maketitle The concept of spin-based electronics demands
semiconductor heterostructures possessing high electron mobility,
pronounced ferromagnetic properties and strong spin-orbit
interaction (SOI).\cite{Dietl_Spintronics,Dyakonov_book} In particular,
manganese doped diluted
magnetic semiconductors (DMS) 
showing
high Curie temperature and large Land\'{e} factor $g^*$
are in the focus of current research.
While enhanced magnetic properties have been obtained in (Cd,MnTe)-
(Ga,Mn)As-based heterostuctures,
the 
SOI in these materials is rather small.
Thus, realization of low-dimensional DMS structures 
based on materials 
which possess 
a strong SOI,
e.g. InAs, becomes important.
Most recently it has been demonstrated that the incorporation of
Mn into a heterostructure device containing an InAlAs/InGaAs
quantum well (QW) leads to a 
two-dimensional hole gas.
\cite{Wurstbauer_2009} In these structures the Mn ions are in
close proximity to the InGaAs channel hosting 
the hole gas.
While DMS hole systems with strong SOI 
have been realized 
and
demonstrate very interesting magnetotransport properties,
\cite{Wurstbauer_2010} the fabrication of InAs-based DMS
with two-dimensional \textit{electron} gas (2DEG) channels has been a
challenge up until now. The 2DEG is characterized by a simple
parabolic band structure and
much higher mobility compared to that of the holes, even in Mn-doped DMS structures like
(Cd,Mn)Te QW; \cite{Furdyna_1988}  features making 2DEG systems attractive for various applications.

Here we introduce a novel concept to grow Mn modulation doped DMS
structures with an InAs 2DEG
channel. 
The structures were fabricated by means of molecular-beam epitaxy (MBE)
growth of III-V/II-VI ``hybrid'' heterostructures with the InAs QW 
following the recipes given 
in Ref.~\onlinecite{Ivanov_2004_1}.
The Mn layers have been inserted into the II-VI barrier 
at distance of 10 monolayers (ML) from InAs QW which is smaller than the magnetic length of 2D electrons.
To explore the magnetic properties of
the 2DEG we investigated spin polarized electric currents induced
by microwave (mw) radiation.\cite{Ganichev_2009, Drexler_2010}
Our measurements show that hybrid AlSb/InAs/(Zn,Mn)Te QW are characterized by 
enhanced magnetic properties which can be changed by
tuning of the spatial position of Mn-doping 
layer as well as by the variation of temperature.

\begin{figure}[htbp]
\includegraphics[width=0.85\linewidth]{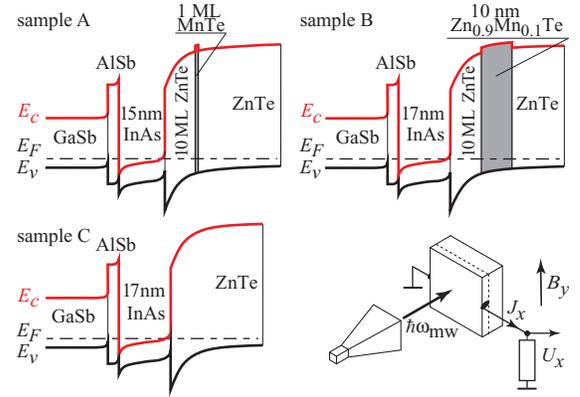}
\caption{Composition and band structure of the investigated
III-V/II-VI 
heterostructures  with an InAs electron 2DEG
channel. Right bottom panel shows the experimental setup.} 
\label{fig1}
\end{figure}

The structures were grown on (001)-oriented GaAs semiinsulating 
substrates at temperature of 280~$^\circ$C.
For the fabrication of AlSb/InAs/(Zn,Mn)Te heterovalent heterostructures 
with different designs of Mn-containing barrier
we used two separated MBE setups. 
The first MBE (Riber 32P, France) was employed
to obtain the III-V part consisting of the 0.2~$\mu $m-thick GaAs and 2~$\mu
$m-thick GaSb buffer layers capped with a 4~nm-thick AlSb barrier and a
15~nm-thick InAs QW (two last layers have common InSb-like interface). 
A (2.5~nm-GaSb/2.5~nm-AlSb)$_{10}$ superlattice was placed
within the first third of the GaSb buffer to suppress propagation of
misfit-induced threading dislocations. 
The II-VI parts of the heterovalent
structures were deposited pseudomorphically on the III-V part in the second
two-chambers MBE setup (Semiteq, Russia) after the ex-situ sulfur chemical
passivation in a 1M Na$_{2} \rm{S} \cdot \rm{9H}_{2}$O solution of the top InAs
layer. The coherent growth of ZnTe on InAs was initiated by simultaneous
opening of Zn and Te fluxes onto a (2x4)As-reconstructured InAs surface
annealed preliminary under an As$_{4}$ flux in the III-Arsenide chamber of
the Semiteq's setup and transferred to the II-VI chamber through ultrahigh
vacuum. Such technology 
results in a high quality coherent AlSb/InAs/ZnTe QW structures 
demonstrating the existence of a 2DEG 
and quantum confined
photoluminescence.\cite{Ivanov_2004_2} According to X-ray
photoelectron spectroscopy 
studies of the InAs/ZnTe 
heterovalent interface grown under similar conditions, its
conduction band offset ($\sim $1.65~eV, see
Ref.~\onlinecite{Gleim_2002}) is close to that for InAs/AlSb
interface.
As a result we obtain AlSb/InAs/ZnTe QW nearly
symmetric for electrons but strongly asymmetric for holes,  which
facilitates  hole escape to the Sb-based barriers even if some
acceptors are introduced into the QW.

To demonstrate that the incorporation of Mn 
leads to enhanced magnetic properties of InAs 2DEG  we prepared
two samples with MnTe insertions (samples \#A and \#B) and one reference structure
of the same design but without Mn (\#C). The structures
have similar composition but differ essentially in the design of
the II-VI barrier grown on top of the QW, as shown in
Fig.~\ref{fig1}. In  sample~A we used 1~ML
thick MnTe insertion separated from the InAs~QW by 10~ML of non-magnetic
ZnTe. Sample~B has the same spacer, but in this structure instead
of 1~ML of MnTe we have grown a 10~nm  of Zn$_{0.9}$Mn$_{0.1}$Te
with substantially smaller concentration of Mn per ML. In the last
reference sample~C no Mn atoms have been added and the II-VI
barrier comprises ZnTe only.

The 2DEG has the density $n \sim $ 1$\times $10$^{13}$\,cm$^{-2}$
and the mobility $\mu \sim 5 \times $10$^3$\,cm$^2$/Vs at a
temperature $T = 4.2$\,K. In all investigated magnetic and
non-magnetic samples
mobilities are almost equal and much smaller than that 
in the AlSb/InAs/AlSb QW of conventional design.
This fact indicates that the electron transport in hybrid structures
is governed mostly by the InAs/ZnTe interface, which seem to
involve fluctuating electrical dipoles due to inhomogeneous distribution of
In-Te and Zn-As bonds and scattering centers of various origin,
while the scattering by Mn atoms is negligible.

The setup used for the  measurements of the mw-induced current 
is shown in Fig.~\ref{fig1}. 
To heat the 2DEG we applied linearly polarized microwave radiation of a
backwardwave oscillator (BWO) operating at frequency
$f$ of 290~GHz or a Gunn diode with $f = 60$~GHz. 
The incident mw power, $P$, of about
2~mW was modulated at 330~Hz by means of a chopper or using a $pin$ switch
for the 
BWO and Gunn diode, respectively.
Unbiased samples of 5$\times $5~mm$^{2}$ with a pair of Ohmic contacts 
centered along opposite  edges were irradiated by mw radiation at 
normal incidence.
The resulted photocurrent $J_{x}$ ($x$~$\vert \vert $~[1$\bar{1}$0]) 
was measured via the voltage drop $U_{x}$ across a 1~M$\Omega$
load resistor applying lock-in technique.

\begin{figure}[htbp]
\includegraphics[width=0.8\linewidth]{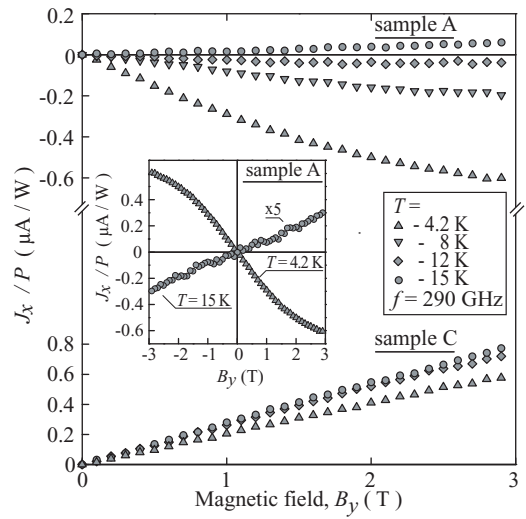}
\caption{Magnetic field dependence of the photocurrent normalized by the radiation power $P$
measured 
for positive $B_y$.  
The inset shows $J_x(B)/P$
for both magnetic field directions.
} 
\label{fig2}
\end{figure}

\begin{figure}[htbp]
\includegraphics[width=0.7\linewidth]{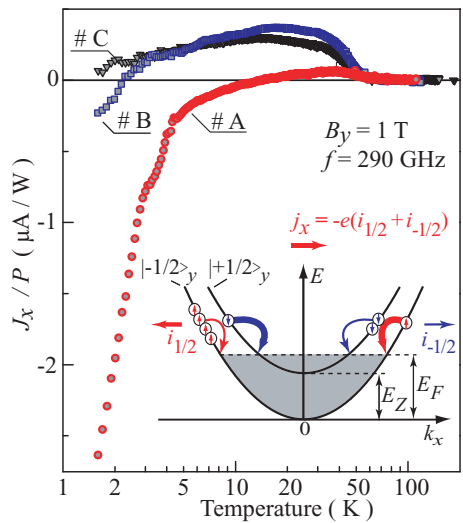}
\caption{
Temperature dependence of the  photocurrent
measured at $B_y = 1$~T. 
The inset shows the model of the mw-radiation induced 
spin polarized electric currents.
Scattering matrix elements linear in
$\bm{k}$ and $\bm{\sigma}$ cause asymmetric scattering probabilities sketched by 
bent arrows of various thickness.
}
\label{fig3}
\end{figure}

Figure~\ref{fig2} shows the  magnetic field dependence of the
photocurrent $J_{x}$ induced in samples~A and~C. The current
increases with $B_{y}$ and reverses its sign as the direction of
$B_{y}$ changes (see inset in Fig.~\ref{fig2}). The temperature
dependence of $J_{x}$ for all samples is plotted in Fig.~\ref{fig3}.
Both figures indicate a remarkable difference in the photocurrents
generated in Mn-doped samples~A and~B and the reference
sample~C. At temperatures above $\sim$60~K the current
in all three structures 
has the same sign and nearly the same magnitude. 
The analogy, however, disappears as the
temperature decreases. In the reference sample~C the
polarity of the signal and its magnetic field behavior remains
unchanged. By contrast, in  structures doped with Mn a reduction of
temperature results in a sign inversion of the current at $T =T_{inv}$.
The most remarkable behavior is detected in sample~A with $T_{inv} \sim 15$~K.  
In this sample below $T_{inv}$ the current increases by more than one order of magnitude 
compared to that 
measured in samples~B and~C at  the same temperature. We also observed that in this sample for 
$T < 8$~K the photocurrent does not depend linearly on $B_y$ anymore, and saturates at high magnetic fields (see Fig.~\ref{fig2}).
In sample~B the inversion temperature is substantially lower ($T_{inv} \sim 2.5$~K) 
and the signal is much smaller than that in sample~A.

All these findings give a strong evidence for enhanced magnetic
properties of the Mn-doped structures and can be well understood
in the frame of the recently proposed model for the spin-dependent
asymmetric energy relaxation of a nonequilibrium 2DEG.
\cite{Ganichev_2009, Drexler_2010}
The inset in Fig.~\ref{fig3} sketches the basic physics of this
phenomenon. Excitation of the 2DEG by mw~radiation causes electron
gas heating. In 
InAs QWs the spin-dependent electron-phonon interaction in 
the energy relaxation 
results in equal and oppositely directed 
electron fluxes, $i_{\pm 1/2}$, for opposite spin subbands,  $|\pm 1/2>_y$.
The application of the 
magnetic field $B_{y}$ emerges a
Zeeman splitting. 
Thus, the electron densities in the subbands
become different, and the fluxes do not 
compensate each other yielding a net electric current $J_{x}$. 
Obviously, the current is spin polarized and its value is
proportional to the Zeeman splitting energy $E_{\rm{Z}}$.

The coupling of the photocurrent sign and magnitude to the band spin
splitting results in a different behavior of $J_{x}$
in 
sample~C and Mn-doped samples~A and~B. In
the non-magnetic structure $E_{\rm Z} = g^* \mu_{\rm B} B$ 
(for bulk InAs $g^* \approx -15$, see Ref.~\onlinecite{Madelung_1982}).
Consequently, the current increases linearly with rising magnetic field
and does not change its sign upon variation of temperature. For
the Mn-doped samples a strong temperature dependence of the band
spin splitting as well as reversing its sign upon temperature
variation is expected if 2D electrons are coupled
to Mn$^{2+ }$ ions. These effects are well known for 2DEG in
(Cd,Mn)Te/(Cd,Mg)Te DMS structures where the band spin splitting
in DMS QWs is given by~\cite{Furdyna_1988}
\begin{equation}
\label{eq1} E_{\rm Z} = g^* \mu_{\rm B} B + x S_0 N_0 \alpha
{\rm B}_{\rm 5/2}\left(\frac{5 \mu_{\rm B} g_{Mn} B}{2k_{\rm B}
(T_{Mn} + T_0)}\right)\,.
\end{equation}
Here $k_{\rm B}$ is the Boltzmann constant,  $\mu _{\rm B}$ the
Bohr magneton, $g_{Mn} = 2$ is Mn $g^*$-factor, 
$T_{Mn} $ is the Mn-spin system temperature, parameters $S_0 $ and
$T_0 $ account for the Mn-Mn antiferromagnetic interaction, $x$ 
is the Mn concentration, $\mbox{B}_{\rm
5/2} \left( \xi \right)$ is the modified Brillouin function, and
$N_0 \alpha $ is the exchange integral.
It is seen that the Zeeman splitting in DMS structures
differs from that in the non-magnetic material by the second term
which is caused by electron exchange interaction with the
Mn$^{2+}$ ions. Equation~(\ref{eq1}) 
explains well the experimental data for our Mn-doped structures.
The effect of the exchange interaction is most pronounced at
low temperatures at which the last term in Eq.~(\ref{eq1}) causes the 
giant Zeeman spin splitting.
The current is proportional to $\mbox{B}_{\rm 5/2} \left( \xi \right)$. Following 
the $E_{\rm Z}$ it is drastically enhanced and saturates at high magnetic fields (see 
Fig.~\ref{fig2}). 
With the temperature increase, the role of the exchange interaction
decreases, $\mbox{B}_{\rm 5/2} \left( \xi \right)$ diminishes,
and, for a certain temperature, the intrinsic band spin splitting
becomes dominant. 
Due to the opposite signs of $g$* and $N_{0}$\textit{$\alpha $} the sign of $E_{\rm Z}$ inverses
resulting in the reversion of the photocurrent direction (see Fig.~\ref{fig3}).\cite{footnote1} 
Lower inversion temperature $T_{inv}$ and substantially smaller
magnitude of the current detected in sample~B in comparison to
that of sample~A indicate the weaker influence of Mn on the magnetic
properties of the InAs 2DEG channel in this sample.

Our data unambiguously demonstrate that 
Mn$^{2+}$ ions crucially affect the magnetic properties of the
InAs 2DEG
channel. In both magnetic samples the Mn doping is done after the InAs 
QW growth (Fig.~\ref{fig1}) and is separated from the QW by rather thick spacer of ZnTe
(10~ML), so that the InAs channel is expected to be free of
manganese.\cite{Wurstbauer_2009,Prechtl_2003} 
The latter is also in agreement with the transport data, because diffusion of Mn into
the InAs channel should yield a hole gas rather then 2DEG.
Thus, we attribute the effect of magnetic ions  to the
exchange interaction caused by the penetration of electronic wave
function 
into the 
barrier.\cite{Meilikhov_2010}
The observed weaker magnetic properties in sample~B comparing to
sample~A supports this mechanism.
In both samples we used Mn insertions with almost the same number
of Mn atoms placed after the 10~ML spacer for which the
magnetic length of 2D electrons remains larger than the spacer in
the whole range of magnetic fields used here. The only difference
is that in sample~A the doping with magnetic ions is localized in
1~ML and in sample~B it is distributed over a larger distance of 10~nm.
Thus, the electron wave function has better overlap with Mn ions
in sample~A compared to sample~B  resulting in increased exchange
interaction and more pronounced magnetic properties.

Summarizing, we show that a structures characterized by a giant Zeeman splitting 
in $n$-type InAs QW and showing behavior typical for DMS
can be obtained by growing of III-V/II-Mn-VI coherent ``hybrid'' heterostructures
with the Mn insertion to the II-VI barrier. 
Our measurements demonstrate that enhanced magnetic properties are due
to penetration of electronic wave function 
into the (Zn,Mn)Te layer and can be controllably varied by
the position and density of Mn$^{2+}$ ions.

The financial support from the DFG (SFB 689 and Priority Group 1483), the
Linkage Grant of IB of BMBF at DLR, RFBR and Russian Ministry of Education
and Sciences is gratefully acknowledged. We are grateful to B. Aronzon and D.~Yakovlev
for fruitful discussions.


\begin{thebibliography}{99}
\bibitem{footnote1} We note, that a similar $J_{x}$ ($T)$ behavior has also been observed recently in (Cd,Mn)Te/(Cd,Mg)Te
QWs where, in contrast to the structures studied here, (Cd,Mn)Te
thin layers were  directly implemented in the QW.
\end{thebibliography}
\end{document}